\documentclass[preprint,showpacs,preprintnumbers,amsmath,amssymb]{revtex4}

\usepackage[english]{babel}
\usepackage{epsfig}
\usepackage{dcolumn}
\usepackage{amsmath}
\usepackage{changebar}
\usepackage{graphicx}

\newcommand{\Tr} {\mbox{Tr}}

\begin{document}
\renewcommand{\figurename}{{\bf Fig.}}
\renewcommand{\tablename}{{\bf Tab.}}
\preprint{BI-TP 2004/28}
\title{The QCD phase diagram: A comparison of lattice and hadron resonance
  gas model calculations}
\author{A.~Tawfik}
 \email{tawfik@physik.uni-bielefeld.de} 
 \affiliation{University of Bielefeld, P.O.~Box
   100131, D-33501~Bielefeld, Germany} 
\date{23. December 2004}

\begin{abstract}
We compare the lattice results on QCD phase diagram for two and three
flavors with the hadron resonance gas model (HRGM) calculations. Lines of
constant energy density $\epsilon$ have been determined at different
baryo-chemical potentials $\mu_B$. For the strangeness chemical potentials
$\mu_S$, we use two models. In one model, we explicitly set $\mu_S=0$ for
all temperatures and baryo-chemical potentials. This assignment is used in
lattice calculations. In the other model, $\mu_S$ is calculated in
dependence on $T$ and $\mu_B$ according to the condition of vanishing
strangeness. We also derive an analytical expression for the dependence of
$T_c$ on $\mu_B/T$ by applying Taylor expansion of $\epsilon$. In both
cases, we compare HRGM results on $T_c-\mu_B$ diagram with the lattice
calculations. The agreement is 
excellent, especially when the trigonometric function of $\epsilon$ 
is truncated up to the same order as done in lattice simulations. For
studying the efficiency of the truncated Taylor expansion, we calculate the
radius of convergence. For zero- and second-order radii, the agreement with
lattice is convincing. Furthermore, we make predictions for QCD phase
diagram for non-truncated expressions and physical masses. These
predictions are to be confirmed by heavy-ion experiments and future lattice
calculations with very small lattice spacing and physical quark masses. 
\end{abstract}


\maketitle

\section{\label{sec:1}Introduction}

The QCD phase diagram at finite temperatures and densities attracted
increasing attention~\cite{Rajagopal:2000wf}, especially as it became
possible to perform lattice QCD simulations at finite baryo-chemical potential
$\mu_B$~\cite{Fodor:2001au,deForcrand:2002ci,Allton:2002zi,D'Elia:2002gd}. 
The numerical studies for the equation of state at finite $\mu_B$ 
provides a valuable framework for understanding the experimental signatures
for the phase transition from confined hadrons to quark-gluon plasma
(QGP). The heavy-ion experiments are aiming to explore the QCD
phase diagram. Therefore, it is of great interest to show the interrelation
between strangeness and baryo-chemical potentials and $T_c$ in hadron
resonance gas model (HRGM) compared with the available lattice QCD
simulations. 

It is known that QCD phase diagram has a very rich structure. From 
numerical simulations, we know that the location of the phase transition
line depends on the quark masses and flavors and the way of
including the strange quark chemical potential $\mu_s$ at different values of
$\mu_B$ and $T$. The isospin chemical potential can play an additional
role. But relative to $\mu_B$ and $\mu_s$, the isospin chemical potential
is very small, so that we can assume an entire symmetry in light quark
potentials and therefore ignore the isospin chemical potential. 

The first point in QCD phase diagram, namely the point at $T_c$ and
$\mu=0$, has been a subject of different lattice
simulations~\cite{Fodor:2001au, deForcrand:2002ci, Allton:2002zi,
  D'Elia:2002gd, Karsch:2000kv, Karsch:2001cy, Gavai:2003mf}. We know 
so far that for two quark flavors ($n_f=2$) the transition is second order or
rapid crossover and the critical temperature is $T_c\approx 173\pm8\;$MeV. For
$n_f=3$, we have a first order phase transition and $T_c\approx
154\pm8\;$MeV. For $n_f=2+1$, i.e. two degenerate light quarks and one
heavy strange quark, the transition is crossover and $T_c\approx
173\pm8\;$MeV. For the pure gauge theory, $T_c\approx 271\pm2\;$~MeV and the
deconfinement phase transition is first order. 

The lattice QCD simulations at $\mu_B\neq0$ is still lacking an effective
exact algorithm and suffer from the sign-problem. The fermion
determinant gets complex and therefore the conventional
Monte~Carlo~techniques are no longer applicable, since the lattice
configurations can no longer be generated with the probability of the
Boltzmann weight. However, during the last few years considerable progress
has been made to overcome these problems~\cite{Fodor:2001au, 
  deForcrand:2002ci, Allton:2002zi, D'Elia:2002gd}.

The significant numerical results on positioning the QCD phase diagram we
have so far is that the transition line can be described by a
parabola. This simple relation can be viewed as a reflection of the truncations
done in the Taylor expansions of different thermodynamic quantities
calculated on lattice. In addition, we know from effective models such as
bootstrap and Nambu-Jona-Lasinio models that the structure of the phase
diagram is complex. The freeze-out curve takes a much different behavior at
large chemical potential~\cite{Tawfik:2004vv, Tawfik:2004ss}. Nevertheless, one
might think that for small chemical potential ($\mu_q\approx T_c$) the
curvature of $T_c$-dependence upon $\mu_q$ can be fitted as a parabola,
where $\mu_q$ is the quark baryo-chemical potential. The situation
at very large chemical potentials is not clear. One might need to take into
account other effects, such as quantum effects at low
temperatures~\cite{Miller:2003ha, Miller:2003hh, Miller:2003ch,
  Miller:2004uc, Hamieh:2004ni, Miller:2004em}, which might be able to
describe the change in the correlations from confined hadrons to 
coupled quark-pairs. 

In present work, we take advantage of our previous
work~\cite{Karsch:2003zq, Karsch:2003vd, Redlich:2004gp} on analysis the
critical temperatures $T_c$ for different quark masses and on reproducing
the lattice thermodynamics at zero and finite $\mu_B$ by HRGM. We assumed
that the deconfinement is driven by a constant energy density. We have
shown~\cite{Karsch:2003zq,Karsch:2003vd} that the degrees of freedom
rapidly increase at $\epsilon_c(T_c,\mu_B=0)$. {\sf Here, we assume that
$\epsilon_c$ remains constant along the whole phase 
transition line~\cite{Redlich:2004gp,Toublan:2004ks},
$\epsilon_c(T_c,0)=\epsilon_c(T_c,\mu_B)$. Concretely, we 
propose that the existence of different transitions does not affect the
assumption that $\epsilon_c$ is constant for all $\mu_B$-values.} We have to
emphasize here that it is not possible in the framework of this model to
make any statement about the transition at very large $\mu_B$ and low
$T$. As $T \rightarrow T_c$, HRGM is no longer applicable. The
nature of the degrees of freedom in this region is 
very different from that of the {\it nearly} non-interacting QGP at high
$T$ and low $\mu_B$. {\sf The message we have in this paper is that the
condition driving the QCD phase transition at finite $T$ and
$\mu_B$~\cite{Karsch:2003zq, Karsch:2003vd, Redlich:2004gp} is the energy
density. Its value is not affected by the conjecture of existing of
different transitions along the whole $\mu_B$-axis}.  

The model will be presented in Sec.~\ref{sec:2a}. In Sec.~\ref{sec:3b}, we
introduce expressions for the lines of constant $T_c$. The lines of
constant energy density are given in Sec.~\ref{sec:3}. The results are
discussed in Sec.~\ref{sec:50}. Sec.~\ref{sec:7} is devoted to the radius
of convergence. In Sec.~\ref{sec:8}, we summarize the conclusions.  
  
\section{\label{sec:2a}The model}

Assuming an ideal quantum gas consisting of point-like hadron resonances,
the canonical partition function for one particle and its
anti-particle reads 
\begin{widetext}
\begin{eqnarray}
{\cal Z}(V,T,\mu) &=& g\frac{V}{2\pi^2} \int_0^{\infty} dk\;k^2
\left\{\ln(1\pm e^{-(\varepsilon-\mu)/T}) + 
\ln(1\pm e^{-(\varepsilon+\mu)/T}) \right\}, \hspace*{10mm} \label{eq:za1}
\end{eqnarray}
\end{widetext}
where $\pm$ stand for bosons and fermions,
respectively. $\varepsilon=(k^2+m^2)^{1/2}$ is the single-particle energy  
and $g$ is the spin-isospin degeneracy factor. Under the given assumptions,
one can sum up the contributions from all resonances pieces, so 
that   
\begin{eqnarray}
\ln {\cal Z}^{(id)}(V,T,\mu) &=& \sum_i^{\infty}  \ln {\cal Z}_i(V,T,\mu_i).
\label{eq:za2} 
\end{eqnarray}
In this expression, there are two important features included; the
kinetic energies and the summation over all degrees of freedom and energies
of resonances. On the other hand, we know that the
formation of resonances can only be materialized through strong
interactions~\cite{Hagedorn:1965st}; {\it Resonances (fireballs) are
  composed of further resonances (fireballs), which in turn consist of
  resonances (fireballs) and so on}.   

In spite of this, if one would like to take into consideration all kinds of
resonance interactions in HRGM, then, for instance, by
means of the $S$-matrix, we can re-write Eq.~(\ref{eq:za2}) as an expansion
of the fugacity term.  
\begin{eqnarray}
\ln {\cal Z}^{(int)}(V,T,\mu) &=& \ln {\cal Z}^{(id)}(V,T,\mu) +
\sum_{\nu=2}^{\infty} a_{\nu}(T) \exp(\mu_{\nu}/T). \label{eq:za3}
\end{eqnarray}
$S$-matrix describes the scattering processes in the thermodynamical
system~\cite{Dashen:1969mb}. $a_{\nu}(T)$ are the so-called virial
coefficients and the subscript $\nu$ refers to the order of the
multiple-particle interactions.
\begin{eqnarray}
a_{\nu}(T) &=& \frac{g_r}{2\pi^3}
\int_{M_{\nu}}^{\infty}dw\;e^{-\varepsilon_r(w)/T}\;
\sum_l(2l+1)\frac{\partial}{\partial w}\delta_l(w). \label{eq:za4}
\end{eqnarray}
The sum runs over the spatial waves. The phase shift $\delta_l(w)$ of two-body
inelastic interactions, for instance, depends on the resonance half-width
$\Gamma_r$, spin and mass of produced resonances,  
\begin{widetext} 
\begin{eqnarray}
\ln {\cal Z}^{(int)}(V,T,\mu) &=& \ln {\cal Z}^{(id)}(V,T,\mu) +
\frac{g_r}{2\pi^3}\int_{M_{\nu}}^{\infty}dw
      \frac{\Gamma_r\;e^{(-\varepsilon_r(w)+\mu_r)/T}
      }{(M_r-w)^2+\left(\frac{\Gamma_r}{2}\right)^2}.
      \hspace*{10mm}\label{eq:za5}  
\end{eqnarray}
\end{widetext}
By inserting $-\mu$ in place of $\mu$ in Eq.~(\ref{eq:za5}), we take into
consideration the two-particle inelastic 
interactions, from which the anti-particles will be produced. For narrow
width and/or at low $T$, the virial term reduces, so that we will get
the {\it non-relativistic} ideal partition function of the hadron
resonances with effective masses $M_{\nu}$. In other words, the resonance
contributions to the partition function are the same as that of free
particles with some effective mass. At temperatures comparable to
$\Gamma_r$, the effective mass approaches the physical one. {\sf Thus, at
  high temperatures, the strong interactions are 
  taken into consideration via including heavy resonances,
Eq.~(\ref{eq:za2}). We therefore suggest to use the canonical partition 
function Eq.~(\ref{eq:za2}) without any corrections.} Furthermore, we do not
apply the excluded volume corrections. We include all hadron
resonances with masses up to $2\;$GeV, such a way we avoid the
singularities expected at Hagedorn
temperature~\cite{Karsch:2003zq, Karsch:2003vd}.  

In the two sections which follow, we discuss how to include the
strangeness chemical potential $\mu_S$ in HRGM and lattice QCD.

\subsection{\label{sec:2}$\mu_s$ in hadron resonance gas model}

\begin{figure}[tb]
\centerline{\includegraphics[width=10.cm]{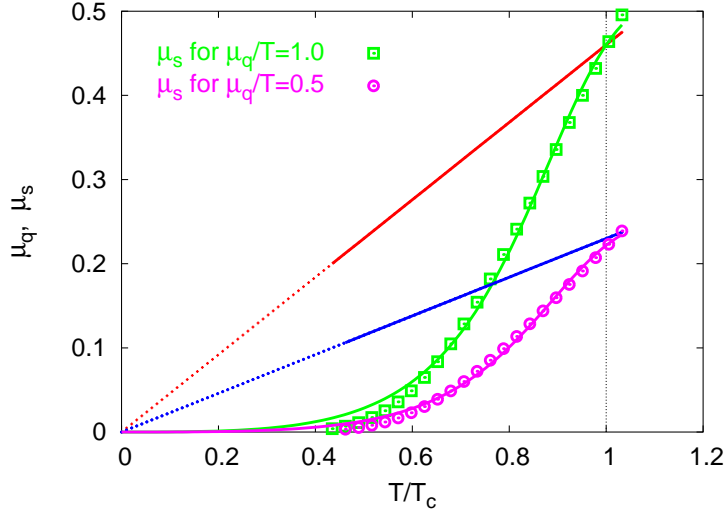}}
\caption{The strange quark chemical potential $\mu_s$ vs. $T/T_c$
  for $\mu_q/T=1$ and $\mu_q/T=0.5$ (straight 
lines). The results are fitted according to Eq.~(\ref{eq:muFit}). At $T=0$,
  we find that $\mu_q=\mu_s=0$. As $T\rightarrow T_c$, the strangeness
  chemical potential approaches the baryo-chemical potential,
  $\mu_s=\mu_q$. The units used here are $\sqrt{\sigma}\sim420\;$MeV.} 
  \label{Fig:1}   
\end{figure}

The {\it hadron}-based chemical potentials $\mu_B,\, \mu_S$ are related to 
the {\it quark}-based ones, $\mu_q,\,\mu_s$ 
\begin{eqnarray}
\mu_{B} &=&   3\,\mu_{q},\hspace*{2cm}  
\mu_{S} =  \mu_{q} - \mu_{s}, \label{eq:mu1}
\end{eqnarray}
Assuming that the isospin and charge chemical potentials are vanishing, we use
the following combination for the hadron resonances 
\begin{eqnarray}
\mu     &=& 3b\mu_{q} + s\mu_{s}, \label{eq:mu2}
\end{eqnarray}
where $b$ and $s$ are the baryon and strange quantum numbers,
respectively. Obviously, this expression is valid for baryons as well as
for mesons. The quantum numbers are entirely conserved.  

The initial conditions in heavy-ion collisions apparently 
include zero net strangeness. This is expected to remain the case during
the whole interaction unless an asymmetry in the production of strange
particles happens during the hadronization. As we are interested in the
hadron thermodynamics and location of QCD phase transition,
we suppose that the net strangeness is entirely vanishing. The average
strange particle number reads    
\begin{eqnarray}
<n_s> &=& \frac{1}{N}\sum_i^N\lambda^{(i)}_s\frac{\partial \ln {\cal
  Z}^{(i)}(V,T,\mu)}{\partial \lambda^{(i)}_s}.
  \label{eq:ns1} 
\end{eqnarray}
$\ln {\cal Z}$ is given in Eq.~(\ref{eq:za3}) and $\mu_S$ is given in
Eq.~(\ref{eq:mu1}). $\lambda_s=\exp(\mu_S/T)$ is the fugacity factor of $s$
quark. The procedure used to calculate the {\it quark}-based $\mu_s$ is the
following: For given $T$ and $\mu_q$ (or $\mu_B$), we iteratively increase
$\mu_s$ and in each iteration, we calculate the difference
$<n_s>-<n_{\bar{s}}>$, Eq.~(\ref{eq:ns1}). The value of $\mu_s$ which
disposes zero net strangeness is the one we read out and shall use in
calculating the thermodynamic quantities. As in Eq.~(\ref{eq:mu1}) the
relation between $\mu_s$ and $\mu_S$ is given by taking into consideration
the baryonic property of $s$ quark. The resulting $\mu_s$ for different
$\mu_q$ (or $\mu_B$) and $T$ are depicted in Fig.~\ref{Fig:1}. 
From this {\it numerical} method, we fit $\mu_s$ as a
function of $T$ and $\mu_B$, 
\begin{eqnarray}
\mu_s & \approx &
  \frac{0.138\;\vartheta\; \theta^3}{1-2.4\;\theta^2
  +2.7\;\theta^3},  \label{eq:muFit} 
\end{eqnarray}
where $\vartheta\equiv\mu_q/T$ and $\theta\equiv T/T_c$.  

Let us note here that for these calculations we have {\it re-scaled} the
resonance masses in order to be comparable to the quark masses
used in lattice QCD simulations. The procedure of giving {\it unphysically}
heavy masses to the hadron resonances is introduced in~\cite{Karsch:2003zq,
  Karsch:2003vd}. In these calculations, the mass of
lightest Goldstone meson becomes \hbox{$m_{\pi}=770\;$MeV} and
consequently, the critical temperature gets almost as large as
$200\;$MeV at $\mu_B=0$. 
 
In order to guarantee vanishing net strange particle numbers in HRGM as the
case in heavy-ion collisions, it is not enough to simply set $\mu_s=0$ and
consequently, \hbox{$\mu_S=\mu_q=\mu_B/3$} in Eq.~(\ref{eq:ns1}). However,
there are publications in which the authors have assigned $\mu_s$ to zero in 
hadron matter and afterwards applied the Gibbs condition for the {\it first
  order} phase transition to QGP. The reason is obvious. Aside the baryons, the
strange mesons with different contents of $s$ quarks play determining
roles at different temperatures and therefore, affect the final results,
Eq.~(\ref{eq:mu1}). Setting $\mu_s=0$, leads to violating the strange
quantum numbers. Nevertheless, we will show here calculations in which we
set $\mu_s=0$. We do this in order to extensively compare with lattice
results. We apply this assignment in order to check the ability of HRGM in
reproducing the current lattice simulations~\cite{Toublan:2004ks}. After
accomplishing this successfully, we can go beyond the lattice constrains to
show the physical picture. {\sf We will make  predictions for QCD phase
  diagram for 
physical masses and non-truncated Taylor series for thermodynamical
expressions. These predictions are to be 
confirmed by heavy-ion experiments and future lattice calculations with
very small lattice spacing and physical quark masses.}  

For completeness of the discussion, we recall the situation in the plasma
regime (see also next section). For conserving strangeness at $T>T_c$, we
have to suppose that $\mu_S=0$. $\mu_S$ consists of one baryonic part
$\mu_B/3$ and another part coming from the strangeness quantum number
$-\mu_s$. From Eq.~(\ref{eq:mu1}), we then get 
\begin{eqnarray}
\mu_s &=& \mu_q=\mu_B/3.
\end{eqnarray}
This result is numerically confirmed in Fig.~\ref{Fig:1}. For $\mu_q=0$ (or
$\mu_B=0$), we find that $\mu_s=0$ for all temperatures. $\mu_s$ increases
with increasing both $\mu_q$ and $T$. At $T_c$, we find that
$\mu_q\approx\mu_s$. Therefore, we can suggest to set $\mu_q=\mu_s$  
for all temperatures above $T_c$. 

We can so far summarize that $\mu_S$ in the hadron matter has to be calculated
in dependence on $\mu_B$ and $T$ under the assumption that the net
strangeness is vanishing. In the QGP phase, one might fulfill this
assumption by setting $\mu_s = \mu_B/3$. As we will see later, in lattice
simulations, one assigns $\mu_S=\mu_s=0$. We deal with all these cases 
in this work.

\subsection{\label{sec:2b}$\mu_s$ in lattice QCD}

In the Euclidian path integral formulation, the partition function of
lattice QCD at finite $T$ and $\mu$ reads 
\begin{eqnarray}
{\cal Z}(T,\mu) &=& \Tr \; e^{-(H-\mu N)/T} \nonumber \\
 &=& \int {\cal D}\psi\;{\cal D}\bar{\psi}\;{\cal D}A\;e^{{\cal
 S}_f(V,T,\mu)+{\cal S}_g(V,T)},  \hspace*{10mm}
\end{eqnarray}
where $(\psi,\bar{\psi})$ and $A$ are the fermion and gauge fields,
respectively. The chemical potential $\mu$ is given in
Eq.~(\ref{eq:mu2}). By Legendre transformation of the Hamiltonian $H$, we
get the Euclidian action \hbox{${\cal S}=\int_0^{1/T}dt \int_V d^3x {\cal
    L}$}. The fermionic action is 
\begin{widetext}
\begin{eqnarray}
{\cal S}_f &=& a^3\sum_x\left[m a \bar{\psi}_x\psi_x + \frac{1}{2}
\sum_{k=1}^{4} \left(\bar{\psi}_x\gamma_{k}\psi_{x+\hat{k}} -
  \bar{\psi}_{x+\hat{k}} \gamma_{k}\psi_x \right)+
  \mu a \bar{\psi}_x\gamma_4\psi_x \right], \hspace*{10mm}
\end{eqnarray}
\end{widetext}
where $a$ is the lattice spacing. As given in Eq.~(\ref{eq:ns1}), the number
density of the quarks with flavor number $x$ is obtained by 
derivation with respect to $\mu_x$,  
\begin{eqnarray}
n_x = \frac{\partial}{\partial \mu_x} \ln {\cal Z}(T,\mu_x). \label{quarkn}
\end{eqnarray}
For checking the dependence of $\mu_s$ on $\mu_q$ and consequently on
$\mu_B$, it is enough to approximate the fermionic part of lattice QCD
Lagrangian for three quark flavors as    
\begin{eqnarray}
{\cal L} &\approx&
\mu_q\left(\sum_{x\in\{u,d\}}\bar{\psi}_x\gamma_4\psi_x  \right) + 
       \mu_s \bar{\psi}_s\gamma_4\psi_s  \nonumber \\
&\approx& \hspace*{2mm} \mu_q n_u + \mu_q n_d + \mu_s n_s. \label{musLatt}
\end{eqnarray}
To taken into account the conservation of the baryon and strange quantum
numbers, the summation in last expression has to run over $s$ quarks,
too. By doing this, last term turns to be $(\mu_q-\mu_s)\,n_s$. Then we
expect that the strangeness on lattice vanishes at $\mu_s=\mu_q$. But from
Eq.~(\ref{musLatt}), which reflects the situation in current lattice
simulations, we find that $n_s=0$ for $\mu_s=0$. 

As discussed in previous section, the strangeness in QGP is 
conserved at $\mu_s=\mu_q=\mu_B/3$. In the hadron regime, especially at large
$\mu_B$, $\mu_s$ (or $\mu_S$) has to be calculated as a function of 
$T$ and $\mu_B$ (Fig.~\ref{Fig:1}). In spite of these considerations, the 
reliable lattice QCD simulations are still limited to \hbox{$\mu_B\approx
 3 T_c$}. As we will see in Sec.~\ref{sec:50}, at this small value, there
is practically no big difference between $\mu_s=0$ and
\hbox{$\mu_s=f(T,\mu_B)$}.

\section{\label{sec:3b}Lines of constant $T_c$}

\subsection{\label{sec:31b}$T_c(\mu)$ in hadron resonance gas
  model}

In Boltzmann limit, the energy density in an ideal quantum system
consisting of one particle and its anti-particle can be expressed as 
\begin{eqnarray}
\epsilon(T,\mu)  &=& \frac{g}{\pi^2} T m^2 \left[m
  K_1\left(\frac{m}{T}\right) + 3 T
  K_2\left(\frac{m}{T}\right)\right]\cosh\left(\frac{\mu}{T}\right).
\label{eq:esp1} 
\end{eqnarray}
$\mu$ is given in Eq.~(\ref{eq:mu2}). We can divide this expression into
two sectors: one meson- $m$ and one baryon-sector
$b$~\cite{Karsch:2003vd,Karsch:2004ti},
\begin{eqnarray}
\varepsilon_h(T,\mu) &=& \varepsilon_m(T) + \varepsilon_b(T)\;
\cosh(\mu/T).  \label{eq:Tayl1}
\end{eqnarray}
Applying truncations in the Taylor expansions of $\cosh(\mu/T)$ up to the
second order of $\mu/T$ - as done in lattice
calculations~\cite{Ejiri:2003dc} - we 
can estimate the lines of constant $T_c$ in HRGM. In doing this, we assume
that \hbox{$\varepsilon_h(T,\mu) = \varepsilon_h(T_c,0)$}. Furthermore, we
assume that the dependence of energy density on the quark mass is quite
small. We see this in Fig.~\ref{Fig:cppacs} which is reported in
Ref.~\cite{AliKhan:2001ek}. At $T_c$ and for the same temporal lattice
dimension $N_{\tau}$, there is very small change in $\varepsilon_c$ for
different quark masses. The latter are related to the pion masses via
$m_{\pi}^2\propto m_q$.    
 
The starting point in expressing $T_c(\mu_B)$ for constant ratios $\mu_B/T$
is to expand Eq.~(\ref{eq:Tayl1}) around the points $T=T_c$ and $\mu_B=0$. The
second-order expansion gives   
\begin{widetext}
\begin{eqnarray}
\varepsilon_h(T,\mu_B) &=& \varepsilon_h(T_c,0) + 
   \left[\frac{\partial \varepsilon_h(T_c,0)}{\partial T}
   (T-T_c) + \frac{1}{2} \mu_B^2 \frac{\partial^2
   \varepsilon_b(T_c,0)}{\partial\mu_B^2} \right],
   \label{eq:eh}\nonumber \\
 &=& \varepsilon_h(T_c,0) + \left[\frac{\partial
   \varepsilon_h(T_c,0)}{\partial T} (T-T_c) + 
   \frac{1}{2}\varepsilon_b(T_c,0)\left(\frac{\mu_B}{T}\right)^2\right].
\end{eqnarray}
\end{widetext}
Under the assumptions given above, this leads to the following
parabola: 
\begin{eqnarray}
\frac{T_c(\mu_B)}{T_c(\mu_B=0)} &=& 1 - \frac{9}{2} \frac{1}{T_c(\mu_B=0)}
\left[\frac{\varepsilon_b(T_c,0)}{\frac{\partial}{\partial T}
    \varepsilon_h(T_c,0)}\right]
\left(\frac{\mu_q}{T}\right)^2, \label{eq:cTc1} 
\end{eqnarray}
where $\mu_q=\mu_B/3$. 
In order to map out the QCD phase diagram by using this analytic
expression, we merely need to calculate the baryon energy density
$\epsilon_b$ at $T_c$ and $\mu_q=0$ and the derivative of the hadron energy
density with respect to $T$, $\partial \varepsilon_h(T_c,0)/\partial
T$. The results are given in Sec.~\ref{sec:50}. Evidently, it is possible
to extend the above expression to include further higher terms of $\mu_q/T$.  

Then to determine $T_c$ at given $\mu_B$, we apply besides the above
analytical method the condition of constant critical energy
density~\cite{Redlich:2004gp,Toublan:2004ks} (see Sec.~\ref{sec:3}). In
this case, {\sf $T_c$ is defined as the temperature at which the energy
  density in HRGM reaches a certain critical value. This value can be taken
  from lattice QCD simulations at $\mu_B=0$}.

\subsection{\label{sec:32b}$T_c(\mu)$ in lattice QCD }

In lattice QCD simulations, the critical temperature $T_c(\mu)$ is to be
calculated from the pseudo-critical coupling $\beta_c(\mu)$ by determining
the susceptibility peak in either the Polyakov loop or the chiral
condensate in $\beta-\mu$ dimensions. The lattice beta function $\beta(a)$
which can be obtained from the string tension is needed in order to
express the results in physical units. From the first non-trivial Taylor
coefficients of $T_c(\mu)$, we get 
\begin{eqnarray} 
\frac{d^2}{d\mu^2}T_c(\mu) &=& -\frac{N_{\tau}^{-2}}{T_c(\mu=0)}
  \frac{\partial^2 \beta_c(\mu)}{\partial \mu^2}
  \left(a^{-1}\frac{\partial a}{\partial \beta}\right), \label{eq:latticeTc}
\end{eqnarray}  
where $N_{\tau}$ is the temporal lattice dimension. 

In the lattice QCD simulations~\cite{Allton:2002zi} with $n_f=2$ and quark
mass $am_q=0.1$, the dependence of $T_c$ on $\mu_q$ has been found to
take the following (parabola) expression: 
\begin{eqnarray} 
\frac{T_c(\mu_q)}{T_c(\mu_q=0)} &=& 1 -
                0.070(35)\left(\frac{\mu_q}{T_c(\mu_q=0)}\right)^2.
                \label{eq:BiSw_Tc1} 
\end{eqnarray}  
In other lattice QCD simulations with the same flavor number but quark
masses four times heavier than the physical
masses~\cite{deForcrand:2002ci},  
\begin{eqnarray} 
\frac{T_c(\mu_q)}{T_c(\mu_q=0)} &=& 1- 0.050(34) 
       \left(\frac{\mu_q}{T_c(\mu_q=0)}\right)^2.  \label{eq:dePh_Tc1} 
\end{eqnarray} 
It has been concluded~\cite{Allton:2002zi} that the last relation
remains almost unchanged for $am_q=0.005$. In this
regard, we have to remember that for small quark masses perturbative
beta function has been used. As we will see later, we actually find an
increase in the curvature with reducing resonance masses from the
values which very well simulate the current lattice calculations
(\hbox{$m_{\pi}\approx 770\;$MeV}) to the physical masses, at which the
lightest Goldstone meson is the physical pion (\hbox{$m_{\pi}=140\;$MeV}). 

The results from Eq.~(\ref{eq:BiSw_Tc1}) are depicted in
Fig.~\ref{Fig:2}. In the same figure, we plot Eq.~(\ref{eq:dePh_Tc1}). We
also plot other lattice results~\cite{Allton:2002zi}, namely the short
vertical lines which give $T_c$ at constant energy density. 

$n_f=3$ lattice QCD results~\cite{Karsch:2003va} for different quark
masses obtained so far can be summarized as 
\begin{widetext}
\begin{eqnarray} 
\frac{T_c(\mu_q)}{T_c(\mu_q=0)} 
        &=& 1 - 0.025(6)\left(\frac{\mu_q}{T_c(\mu_q=0)}\right)^2,
        \hspace*{10mm} a m_q=0.1,  \label{eq:BiSw_Tc2} \\ 
        &=& 1 - 0.114(46)\left(\frac{\mu_q}{T_c(\mu_q=0)}\right)^2,
        \hspace*{8mm} a m_q=0.005. \label{eq:BiSw_Tc3}   
\end{eqnarray}  
\end{widetext}
Again, in these simulations the beta function for the small quark masses has
been calculated perturbatively. The results from these two expressions are
depicted in Fig.~\ref{Fig:3}~and~Fig.~\ref{Fig:3b}, respectively. 
From other $n_f=3$ lattice QCD simulations~\cite{deForcrand:2003hx}, 
\begin{eqnarray} 
\frac{T_c(\mu_q)}{T_c(\mu_q=0)} &=& 1- 0.0610(90) 
       \left(\frac{\mu_q}{T}\right)^2 + 0.00235(89) 
       \left(\frac{\mu_q}{T}\right)^4. \hspace*{6mm} \label{eq:dePh2_Tc}
\end{eqnarray}
The curvature from $n_f=2+1$ lattice simulations~\cite{Fodor:2002sd} has
been compared with the three flavor one~\cite{deForcrand:2002ci} and
concluded that there is a complete agreement. As we
will see in Fig.~\ref{Fig:3b}, the most recent $n_f=2+1$ lattice
simulations of~\cite{Fodor:2004nz} result in a curvature smaller
than those from other lattice simulations. This can be understood according
to the different fermionic actions used.

\section{\label{sec:3}Lines of constant energy density}

In previous work~\cite{Redlich:2004gp}, we have used HRGM in order to
determine $T_c$ corresponding to a wide range of quark (pion)
masses at $\mu_B=0$. The masses range from the
chiral to pure gauge limits. We have seen that the condition of 
constant energy density can excellently reproduce the critical temperature
$T_c$ as a function of $m_q$ and $n_f$.  
In present work, we extend this condition to finite chemical
potentials $\mu_B$. We use two models for including the strangeness chemical
potential $\mu_S$. In the first model, we calculate $\mu_S$ in dependence
on $\mu_B$ and $T$ according to the condition of zero net strangeness. In
the other one, we explicitly set the quark-{\it based} $\mu_s=0$. The
last case is used in lattice QCD simulations. As given in
Sec.~\ref{sec:2b}, the proper inclusion of $\mu_s$ in QGP and in lattice
Lagrangian is to assign to it the same value of $\mu_B/3$. Nevertheless,
both choices can be accepted for the current lattice QCD simulations, since
the most reliable lattice calculations are currently performed at small
baryo-chemical potentials (\hbox{$\mu_q\approx T_c$}). Consequently, the
strangeness chemical potential is expected to be much smaller or at lest as
small as the baryo-chemical potential.   

As discussed above, in HRGM, the energy density at finite chemical
potential can be divided into two parts: one from the meson sector and
another one from the baryon sector. For the first part, we can completely
drop out the fugacity term. For symmetric numbers of light quarks, the
baryo-chemical potential of mesons is vanishing. But for strange mesons the
strangeness chemical potential assigned to their $s$ quarks should be taken
into account. For the baryon sector, the chemical potential is given by
Eq.~(\ref{eq:mu2}).

\begin{figure}[tb]
\includegraphics[width=10.cm]{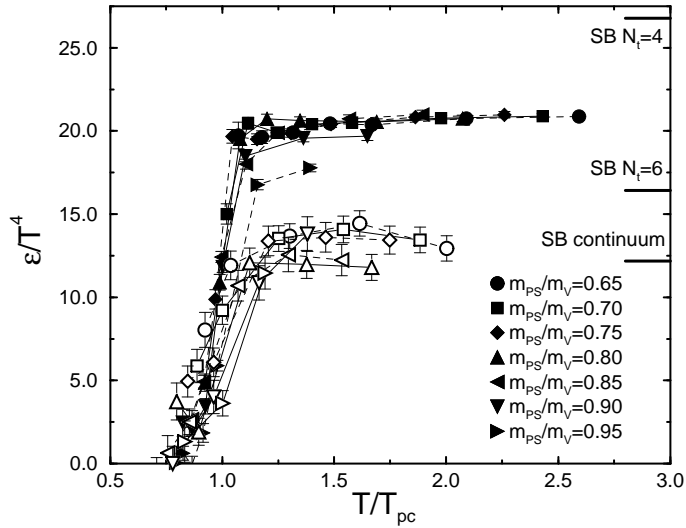}\vspace*{-.9cm}
\caption{\footnotesize The energy density normalized to $T^4$ given in
  dependence on 
  $T/T_c$ for different quark masses $m_q$ and temporal lattice dimensions
  $N_{\tau}$. This figure is reported in Ref.~\cite{AliKhan:2001ek}. } 
\label{Fig:cppacs}
\end{figure}

The question we intent to answer is: which value has to be assigned to the
critical energy density at finite chemical potentials? We recall the
lattice QCD simulations. In Fig.~\ref{Fig:cppacs}, we see that the energy
density~\footnote{Comparing full QCD with pure gauge results, we get a
feeling about the dependence of critical values on quark
mass. $\epsilon_c/T^4$ seem to be different. On the other hand, taking into
consideration the different critical temperatures, we find that the
$\epsilon_c$ are comparable with each other.} at $T_c$ is not a singular
function but can rather be defined at different {\it critical} values
depending on $n_f$ and $m_q$. This reflects the nature of the phase
transition, {\it cross-over}. On the other hand, the uncertainty in
calculating the energy density on lattice is very large, almost a factor
$2$. The reason for this is the uncertainty in estimating $T_c$,
Eq.~(\ref{eq:BiSw_Tc1}). As mentioned above, $T_c$ is determined according
to maximum susceptibility. The uncertainty in this quasi $T_c$ is $\sim
10\%$. The coefficients of $T$ have additional $\sim 20\%$. 

The critical energy density at which we define $T_c(\mu_B)$ in HRGM is
taken from lattice QCD simulations at $\mu_B=0$~\cite{Karsch:2000kv}. In
lattice units, the dimensionless energy densities for $n_f=2$ and $2+1$ are
\hbox{$\varepsilon/T^4|_{T_c}\cong 4.5\pm2$} and $\cong 6.5\pm2$,
respectively. We take an average value and express it in physical
units. Thus, $\epsilon_c=600\pm300\;$MeV$/$fm$^3$. 

As in~\cite{Allton:2002zi}, we assume that this value remains
constant along the phase transition line; $\mu_B$-axis. The existence of
different phase transitions (cross-over and first-order) and the critical
endpoint, at which the transition is second-order, is assumed not to affect
this assumption. As mentioned in Sect.~\ref{sec:32b}, the phase transition
line on lattice is to be estimated according Eq.~(\ref{eq:latticeTc}),
i.e. up to $(\mu/T)^2$. Additional to this method, the line of constant
energy density has been calculated~\cite{Allton:2002zi}. By comparing the
two lines, it has been found that they are almost coincident. In other
words, the constant $\epsilon_c$ can be used to estimate the
critical line at finite $\mu_B$. We have to remember that the lattice
estimation of $\epsilon$ has a very large uncertainty, $\epsilon\propto
T^4$. Therefore, the transition line according to constant $\epsilon_c$ has
a larger uncertainty, because of the additional uncertainty in the
derivatives of $\epsilon_c$ with respect to $\mu_B$ and $\mu_B^2$.

What are the consequences if the assumption of constant $\epsilon_c(\mu_B)$ 
disregarding the uncertainty turned out to be incorrect? To discuss
such consequences, we first recall the second-law of thermodynamics.
\begin{eqnarray}
\partial \epsilon &=& T\,\partial s - p + \mu_B\,\partial n_B, \label{2law}
\end{eqnarray}
where $s$, $p$ and $n_B$ are the entropy, the pressure and the
baryon number density, respectively. 

If $\epsilon_c$ were a decreasing function of $\mu_B$, this means that low
incident energies are much more suitable to produce the phase
transition from confined hadrons to deconfined QGP than the high incident
energies! But the low limit should be given by the freeze-out
curve~\cite{Tawfik:2004ss}, i.e., the hadronization phase diagram. We know
from phenomenological observations that both freeze-out curve and phase
transition line are coincident at low $\mu_B$ (high incident energy). For
large $\mu_B$ the two lines are separated.  The freeze-out
curve~\cite{Tawfik:2004ss} is given by $s/T^3=7$, entropy driven. Then
along freeze-out curve it is expected that $\epsilon$ slightly increase
with $\mu_B$. This means the assumption that $\epsilon_c$ decreases with
increasing $\mu_B$ will give a phase transition which has $T_c$ much
smaller than the freeze-out temperature.

In the other case, that $\epsilon_c$ increases with increasing $\mu_B$, we
expect the $\epsilon_c$ required for the phase transition gets larger with
decreasing the incident energy! According to Eq.~(\ref{2law}), the phase
diagram is given by $T_c=\partial \epsilon/\partial s$ at constant
$n_B$ and $\mu_B= \partial \epsilon/\partial n_B$ at constant $T$. Then for
increasing $\epsilon_c(\mu_B)$, the critical temperature is expected to
increase or at least remain constant. This result has the conconsequences
that the phase transition at large $\mu_B$ will be diffecult to
materialize in heavy-ion collisions or impossible. Also the phases of
coupled quark-pairs (color superconductivity) will be expected for very
large $\mu_B$ or not allowed at all.    

Since we use constant $\epsilon_c$ down to low temperatures, one might
ask if there could be any relation between $\epsilon_c$ and the experimental
value of nuclear density. Obvioulsy at $T=0$, HRGM, Eq.~(\ref{eq:za1}), is
no longer applicable. In this limit, we suppose that the hadron gas is
composed of degenerate Fermi gas of nucleons. We can therefore calculate
$\mu_B$ corresponding to the normal nuclear density at
$T=0$~\cite{Tawfik:2004ss}. The value is $979\,$MeV.  The energy density in
this limit reads 
\begin{eqnarray}
\epsilon(\mu_B) &=& \frac{3\,g}{4\pi^2}\, m^4\;\left[
       \frac{\mu_B}{m}  
       \sqrt{\frac{\mu_B^2}{m^2}-1}
       \left(\frac{\mu_B^2}{m^2}-\frac{1}{2}\right) - \frac{
       \ln\left(\frac{\mu_B}{m} +
       \sqrt{\frac{\mu_B^2}{m^2}-1}\right)}{2}\right]
\label{eq:raioT0}
\end{eqnarray}

\section{\label{sec:50}The results}

Since we want to compare HRGM with lattice results~\cite{deForcrand:2002ci,
Allton:2002zi}, some details about the lattice simulations at
\hbox{$\mu_q\neq0$} are in order. Extensive details about HRGM are
presented in Refs.~\cite{Karsch:2003zq, Karsch:2003vd, Redlich:2004gp}. In
order to avoid the sign problem in lattice
simulations~\cite{Allton:2002zi}, the  derivatives of thermodynamic
quantities with respect to \hbox{$\mu_B=0$} at the point $T_c$ are first
computed and then their Taylor expansion coefficients in terms of finite
$\mu_q$ are calculated. The curvatures according to $T_c(\mu_B)$ and  
$\epsilon_c(\mu_B)$ are derived for different $m_q$. For instance
for $am_q=0.1$, the corresponding pion mass in lattice units for
$n_f=2$ is $a m_{\pi}=0.958(2)$, where
$a\sigma=0.271(10)$~\cite{Karsch:2000kv}. This leads to $m_{\pi}\cong
773\;$MeV. Keeping these features in mind, we performed our calculations
for physical and re-scaled resonance masses. In  Fig.~\ref{Fig:2}, we find
that our results are coincide with the lattice ones. In
Ref.~\cite{deForcrand:2002ci}, the lattice simulations are performed with
an imaginary chemical potential for $n_f=2$ staggered quarks. For
$\mu=i\mu_I$, the sign problem is obviously no longer existing. Using
analytic continuation of the truncated Taylor series, one  can then go to
real $\mu$. The light quark mass is four times the physical mass. The
location of $T_c$ at $\mu_B=0$ has been extrapolated to the chiral
limit. The lattice results are represented by  the short curves in
Figs.~\ref{Fig:2},~\ref{Fig:2b}~and~\ref{Fig:3b}.

\subsection{\label{sec:5}Results for two flavors}

From the canonical partition function Eq.~(\ref{eq:za1}), we can derive
the energy density at finite chemical potential $\mu\neq0$ as 
\begin{eqnarray}
\epsilon(T,\mu) &=& T \frac{\partial T\ln {\cal Z}(T,\mu)}{\partial T} -
  T \ln {\cal Z}(T,\mu) 
  +\mu \frac{\partial T\ln {\cal Z}(T,\mu)}{\partial
  \mu} \nonumber \hspace*{5mm} \\
  &=& \frac{g}{2\pi^2}\; \int_{0}^{\infty} k^2 \;dk \;
  \frac{\varepsilon(k)}{e^{[\varepsilon(k)-\mu]/T} 
  \pm 1}. \hspace*{8mm} 
\label{eq:epslCom}
\end{eqnarray}
In Boltzmann limit and by taking into consideration only one
particle and its anti-particle, we get the expression given in
Eq.~(\ref{eq:esp1}). It is obvious that the trigonometric function included  
in last expression are not truncated. In calculating this quantity in HRGM,
we sum up over all resonances we take into account.  

We first start with two flavors. It is not needed to care about the
strangeness chemical potential. As discussed above, there are two lattice
QCD simulations for $n_f=2$. In the first one, re-weighting methods, the
physical quantities which usually can be calculated at $\mu_B=0$ without
any difficulties have responses at finite $\mu_B$. The responses are
utilized to estimate the thermodynamic quantities at finite $\mu_B$. The
other lattice simulations~\cite{deForcrand:2002ci} use imaginary $\mu_B$
and afterwards apply analytical continuation. At given $\mu_B$, $T_c$ is
determined as the temperature at which the energy density $\epsilon$
equals $600\;$MeV$/$fm$^3$.   
 
\begin{figure}[tb]
\includegraphics[width=10.cm]{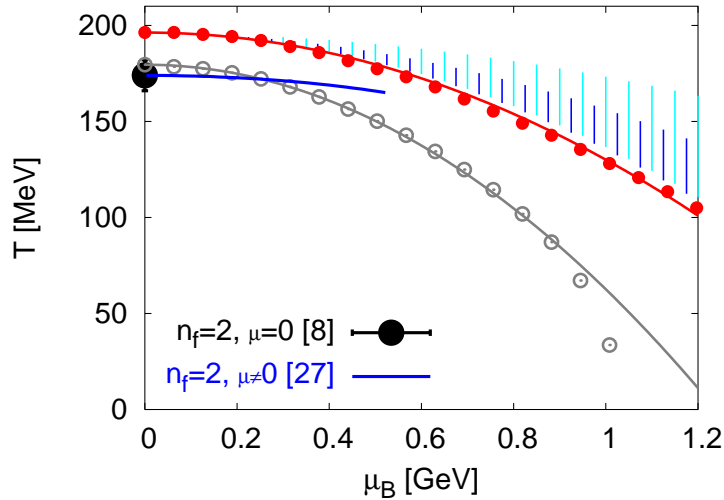} 
\caption{\footnotesize $T-\mu_B$ phase diagram for two quark flavors
  $n_f=2$. The vertical lines give the lattice
  results~\cite{Allton:2002zi}. The short lines show the results according
  to a constant $\epsilon_c$. The long ones are for constant $T_c$. The
  Lattice simulations are performed for large quark mass. The 
  corresponding Goldstone pion gets a mass of $770\;$MeV. The solid circles
  give our results heavy quark masses. The results for
  physical quark masses are given by the open circles. 
}
\label{Fig:2}
\end{figure}

\begin{figure}[tb]
\includegraphics[width=10.cm]{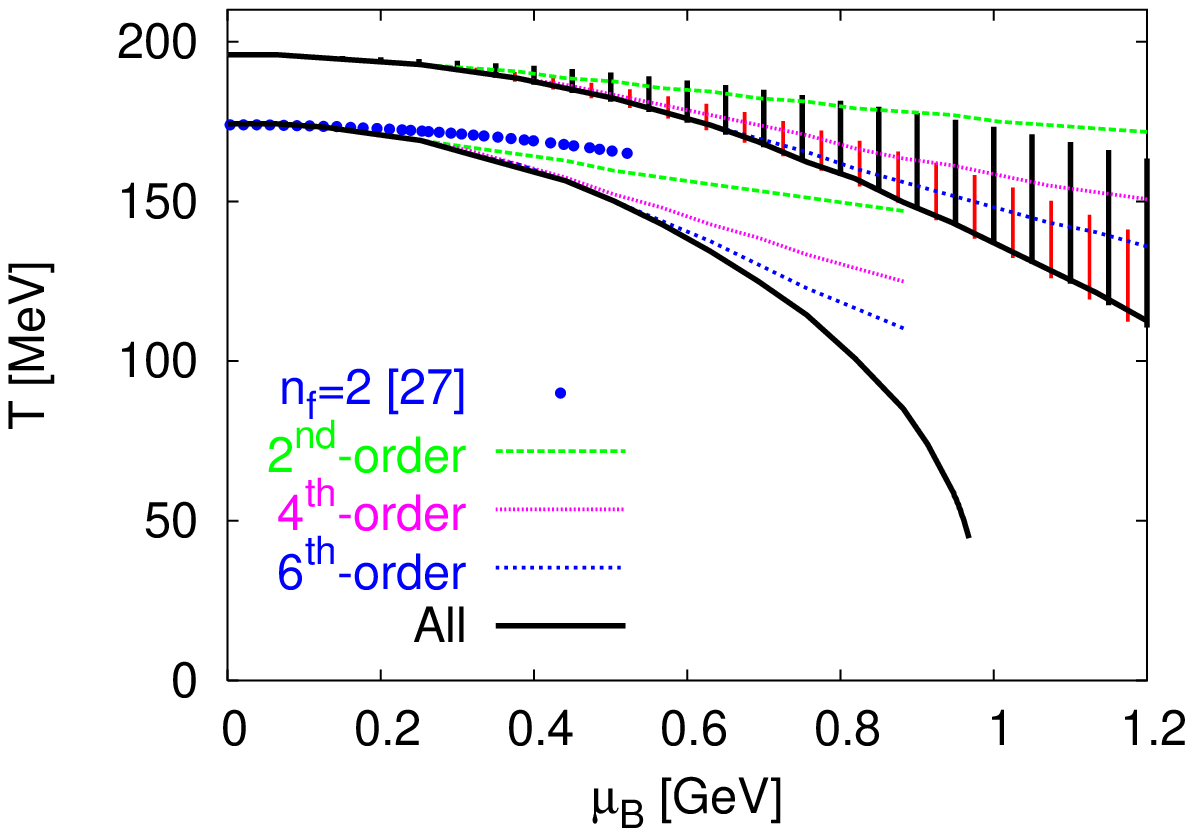}
\caption{\footnotesize The same as in Fig.~\ref{Fig:2}. Here, we show 
  the results obtained from the condition of constant $\epsilon_c$
  truncated up the different Taylor orders. The truncation is obviously
  able to describe the lattice QCD simulations for different quark
  masses~\cite{Allton:2002zi,deForcrand:2003hx}. The solid lines give our
  predictions for non-truncated trigonometric functions. } 
\label{Fig:2b}
\end{figure}

The results are plotted in Fig.~\ref{Fig:2}. The solid circles represent
the results for {\it re-scaled} resonance masses. The open circles 
represent the results for physical masses. The two lines connecting the
points are obtained by fits according to: 
\begin{eqnarray}
\frac{T_c(\mu_q)}{T_c(\mu_q=0)}  &=& 1 - c_1
              \left(\frac{\mu_q}{T_c(\mu_q=0)}\right)^2. \label{eq:FitParab}
\end{eqnarray}
The $\mu_q$-values are restricted within the range $0\leq\mu_q\leq
T_c$. $\mu_q$ is given in Eq.~(\ref{eq:mu1}). For
the re-scaled heavy masses, the fit parameters are \hbox{$c_1=0.115(36)$} and
\hbox{$T_c(\mu_B=0)=196.3\;$MeV}. Plugging these parameters
in Eq.~(\ref{eq:FitParab}), we get the top line in Fig.~\ref{Fig:2}. The
comparison with the lattice calculations Eq.~(\ref{eq:BiSw_Tc1})
gives a satisfactory agreement, especially at low $\mu_B$. Nevertheless, it is 
obvious that our results at large $\mu_B$ lie below the lattice ones. The
reason for this discrepancy will be discussed later.  

One might ask whether the pion gas with very heavy masses would be able to
describe the lattice results? To answer this question, we refer
to~\cite{Karsch:2003vd, Karsch:2003zq}. In order to simulate the lattice
QCD thermodynamiscs, for instance the rapid increase in $\epsilon$ near
$T_c$, we need to include the corresponding degrees of freedom in 
HRGM. A hadron gas of pions can reproduce $\sim 15\%$ of the lattice QCD
thermodynamiscs. Including heavier resonances gives the correct QCD
thermodynamics. Furthermore, for lattice calculations with very heavy quark
masses (pure gauge) we need to include the {\it low-lying}
glueballs~\cite{Karsch:2003vd}.   

As given above, the coefficient in the front of $(\mu_q/T)^2$ in
Eq.~(\ref{eq:cTc1}) can be calculated in HRGM. For non-strange hadron
resonances with re-scaled masses, the coefficient gets the value
$0.0767$. Comparing to Eq.~(\ref{eq:BiSw_Tc1}), this value is obviously
much better than the value of $c_1$ in describing the lattice results. As
we will see, this discrepancy is to be related to the fact that $c_1$ has been
calculated from fitting $T_c$ at finite $\mu_B$ in the results which we have
obtained from non-truncated $\epsilon(T,\mu_B)$. We should emphasize that all 
Taylor terms that are higher than the second one are explicitly excluded in
deriving Eq.~(\ref{eq:cTc1}). This partially explains the disagreement
between HRGM and lattice results shown so far. We also
study the case for physical resonance masses. The fit parameters are
\hbox{$c_1=0.195(21)$} and \hbox{$T_c(\mu=0)=175\;$MeV}. The coefficient in
Eq.~(\ref{eq:cTc1}) is found to be $0.1368$.


In the following, we confront the lattice results with HRGM results which
have been obtained from truncated expressions. We show the results in
Fig.~\ref{Fig:2b}.  We start
with results from entire Taylor expansion for heavy resonance masses
(solid line). This is equivalent to use Eq.~(\ref{eq:esp1}) instead of
Eq.~(\ref{eq:epslCom}). We note that this line shows almost the same
behavior as that of the solid circles in Fig.~\ref{Fig:2}. In deriving
Eq.~(\ref{eq:esp1}), the Boltzmann limit is assumed. The results from
different truncations are also plotted. We find that the energy density
truncated up to the second or fourth order of $\mu/T$ can describe the
lattice results better than the solid line. The agreement between the
curvatures of these two lines and that from the analytical expression
Eq.~(\ref{eq:cTc1}) is excellent. The ability of truncated expression
to produce results comparable to the lattice ones is to be explained by the
fact that the lattice results themselves have been obtained from truncated
thermodynamic expressions.  
 
We plot in the same figure the HRGM results with physical masses. {\sf The
curves obtained from different truncation terms in Eq.~(\ref{eq:esp1})
represent our predictions when it will be possible to perform lattice
simulations for two quark flavors with physical masses. These predictions
has to be checked by future lattice simulations.} We note that the solid
line is comparable to the bottom points plotted in Fig.~\ref{Fig:2}. The
quark masses used in Ref.~\cite{deForcrand:2003hx} are relative
heavy. Nevertheless, we see that the second order (dashed line) agrees very
well with these lattice results~\cite{deForcrand:2003hx} (solid circles).

\subsection{\label{sec:6}Results for three flavors}

\begin{figure}[tb]
\includegraphics[width=10.cm]{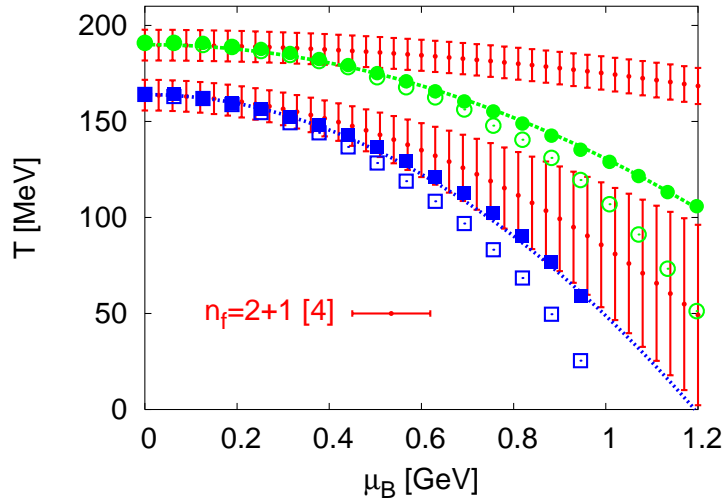}
\caption{\footnotesize The $T-\mu_B$ phase diagram for $n_f=3$. The
  vertical short lines represent the lattice
  results~\cite{Allton:2002zi}. The above band corresponds to heavy quark 
  masses. The bottom one is for physical masses. The solid circles
  are our results for heavy resonances and for $\mu_s=0$. The open circles
  are for $\mu_s=f(\mu_B,T)$. The squares give the results for the physical
  masses. The curves are fitted according to Eq.~(\ref{eq:FitParab}). 
} 
\label{Fig:3}
\end{figure}

We have to include strangeness chemical potential $\mu_S$
(Sec.~\ref{sec:2}). Assuming that the three quarks are 
degenerate, we can use the same critical energy density  as done
in previous section. The results are shown in Fig.~\ref{Fig:3}. The
solid circles show the results for {\it re-scaled} heavy resonance
masses. As done in lattice calculations, we first set $\mu_s=0$. The resulting 
points (solid circles) are fitted within the range \hbox{$0\leq\mu_q\leq
 T_c$} according to Eq.~(\ref{eq:FitParab}). The fit parameters are
\hbox{$T_c=190\;$MeV} and \hbox{$c_1=0.1016$}. The coefficient in front of
$(\mu/T)^2$ in Eq.~(\ref{eq:cTc1}) is $0.0505$. Comparing with
Eq.~(\ref{eq:BiSw_Tc2}), this value can describe the lattice curvature much
better than $c_1$. We also calculate $\mu_s$ in dependence on $\mu_B$ and $T$
assuming strangeness conservation. The open circles show the $T_c-\mu_B$
diagram corresponding to this value of $\mu_s$. We
note that the former case ($\mu_s=0$) is much closer to the lattice results
than the latter one. The reason is, as we mentioned above, that $\mu_s$ in
lattice calculations used to be assigned to zero.   

The results for physical masses are also shown in Fig.~\ref{Fig:3}. The
solid squares represent the results at $\mu_s=0$. These points are fitted
according Eq.~(\ref{eq:FitParab}). The fit parameters are $T_c=164\;$MeV
and $c_1=0.17$. The coefficient in Eq.~(\ref{eq:cTc1}) takes the value
$0.1122$, which agrees very well with Eq.~(\ref{eq:BiSw_Tc3}). The open
squares represent the results at $\mu_s$ that depends on $\mu_B$ and
$T$. The results with {\it re-scaled} resonance masses are given as 
circles. For $\mu_s=0$, we get results closer to lattice results than 
for $\mu_s=f(T,\mu_B)$.
In this figure, the energy density is calculated according to
Eq.~(\ref{eq:epslCom}). In other words, the results plotted here are
deduced from expression like Eq.~(\ref{eq:esp1}), i.e. without any
truncations.    

\begin{figure}[tb]
\centerline{\includegraphics[width=10.cm]{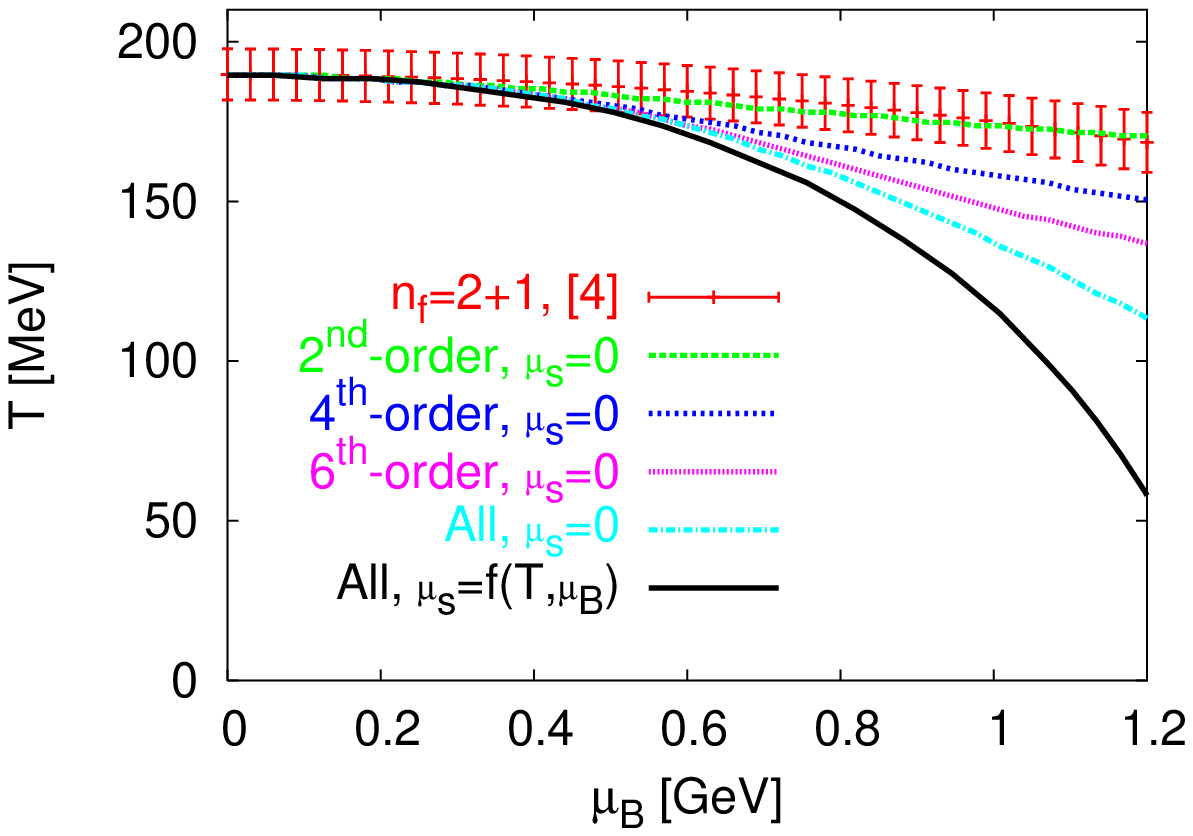}} 
\centerline{\includegraphics[width=10.cm]{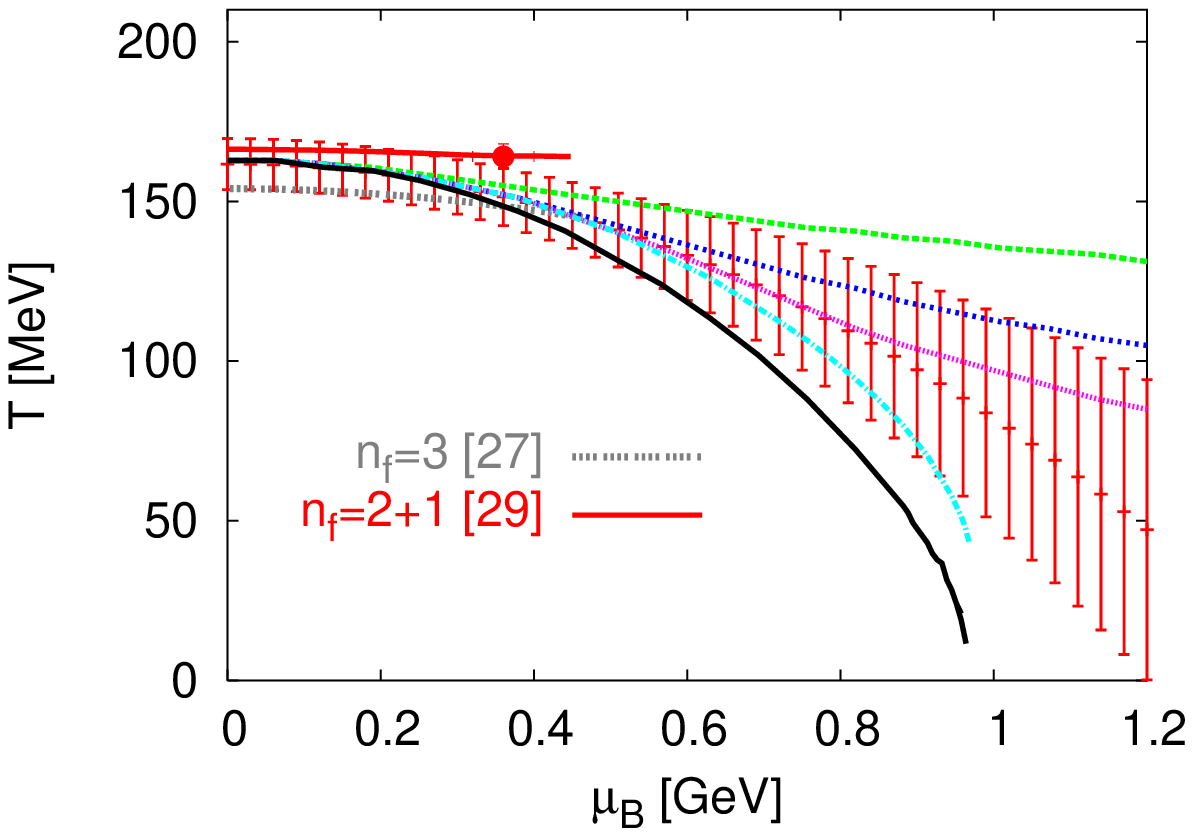}}
\caption{\footnotesize $T-\mu_B$ phase diagram as in
  Fig.~\ref{Fig:3}. Here, we check the effects of truncated trigonometric
  functions. The top panel shows the results for heavy
  masses. Truncating $\epsilon(T,\mu_B)$ up to  
  second order reproduces the lattice results~\cite{Allton:2002zi}. In
  the bottom panel we show the results with physical masses. the lattice
  results~\cite{Allton:2002zi} are very well reproduced by the condition of
  constant truncated $\epsilon$. The results from non-truncated $\epsilon$ at
  $\mu_s=0$ and $\mu_s=f(T,\mu_B)$ are also plotted, the bottom curves,
  respectively. The agreement with the
  lattice simulations~\cite{deForcrand:2003hx,Fodor:2004nz} is also
  convincing } 
\label{Fig:3b}
\end{figure}

We plot in Fig.~\ref{Fig:3b} the results from different truncations, in
order to compare HRGM results with lattice simulations. For heavy masses
the results are given in the top panel. The results for physical masses are
given in the bottom panel. {\sf It is 
clear that the truncation up to the second order gives results in a good
agreement with the lattice simulations~\cite{Allton:2002zi}.} With higher
truncations, we get curves with larger curvatures. We also plot two curves
from non-truncated expressions for the trigonometric functions. The solid
curve is obtained by computing 
$\mu_s$ in dependence on $T$ and $\mu_B$. Obviously, it is identical to the
solid symbols plotted in Fig.~\ref{Fig:3}. The dashed curve - next to the
solid one - shows the results in which $\mu_s$ is entirely vanishing. It
gives the same behavior as that of the open symbols in
Fig.~\ref{Fig:3}. 

{\sf We can so far conclude that the lattice results~\cite{Allton:2002zi} can
be reproduced by HRGM, if the trigonometric functions are truncated
in the same way. The lattice curvature within the region
$0\leq\mu_q\leq T_c$ can excellently be described by HRGM.  }

In the bottom panel, we show other three flavor lattice results. 
In Ref.~\cite{deForcrand:2003hx}, the numerical simulations have been
performed with Wilson gauge action and three degenerate flavors of
staggered fermions. The quark masses are ranging between \hbox{$0.025\leq a
  m_q\leq 0.04$}. It is clear that our results can reproduce the structure
given by these
data. We see that the curvature can be described by second or fourth
order in HRGM. This is also valid for \hbox{$n_f=2+1$} lattice
results~\cite{Fodor:2004nz}. In this case, the quark masses
\hbox{$am_{u,d}=0.0092$} and \hbox{$am_{s}=0.25$}. Chiral 
extrapolations are done by heavier light quark masses. 
We also plot the latest calculations of the location of the endpoint. 
The endpoint in~\cite{Allton:2003vx} lies at \hbox{$\mu_B=420\;$MeV}. The
corresponding temperature has not yet been calculated. The endpoint
in~\cite{Fodor:2004nz} has the coordinates \hbox{$\mu_B=360\pm40\;$MeV} and
\hbox{$T=162\pm2\;$MeV}. As discussed above, we assume that the
existence of endpoint does not affect our results.

\section{\label{sec:7}Radius of convergence}

\begin{figure}[thb]
\includegraphics[width=10.cm]{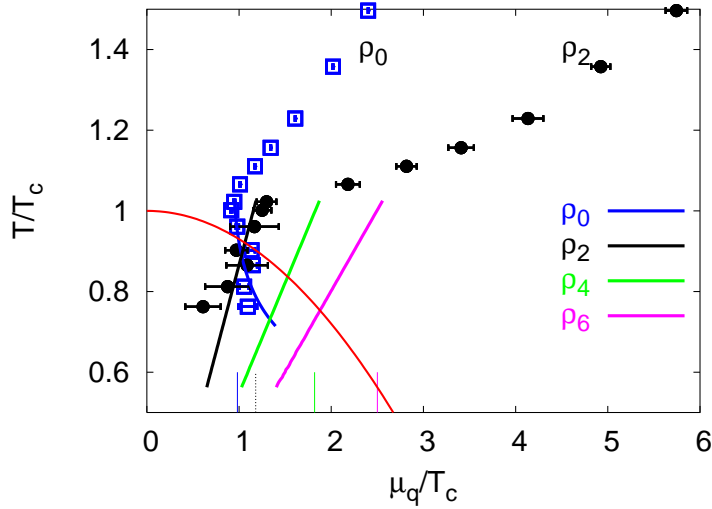}
\caption{\footnotesize Comparison between radii of convergence
  from the lattice~\cite{Allton:2003vx} (points) and HRGM
  (lines). There is an excellent agreement with the existing lattice
  results. The $T_c$-curvature has been calculated by using lattice results
  with two flavors (Fig.~\ref{Fig:2}). The short vertical lines give the
  values of 
  corresponding radii at $T_c$.} 
\label{Fig:5}
\end{figure}

As we have seen, for a reliable comparison with the current lattice
simulations, we have to apply truncations in the thermodynamic expressions
in HRGM. Our objective here is to check the efficiency of truncated
series in locating the phase diagram. The radius of convergence reflects the
singularity near $T_c$. It approaches unity near $T_c$. 
The energy density normalized to $T^4$ can be expressed as a trigonometric
function depending on the ratio $\mu/T$~\cite{Redlich:2004gp}
(Sec.~\ref{sec:3b}). We use the property that the energy density is an even
function in $\mu_B/T$ and therefore write  
\begin{widetext}
\begin{eqnarray}
\frac{\epsilon(T,\mu_B)}{T^4} &=& \epsilon_m(T) + \epsilon_b(T)
  \cosh\left(\frac{\mu_B}{T}\right) \nonumber \\ 
  &\approx&  \epsilon_b(T)\left[c_2\left(\frac{\mu_q}{T}\right)^2
  + c_4\left(\frac{\mu_q}{T}\right)^4 \cdots \right], \label{eq:pexpan1} 
\end{eqnarray}
\end{widetext}
where $c_n=(T_c/T)^n\;3^n/n!$ and $n$ is an even positive
integer. The radius of convergence of Taylor expansion of the partition
function, Eq.~(\ref{eq:esp1}) is to be estimated by the
ratios of subsequent expantion coefficients,
\begin{eqnarray}
\rho &=& \lim_{n\rightarrow\infty}\left|\frac{c_n}{c_{n+2}}\right|^{1/2}. 
\label{eq:rhon}
\end{eqnarray}
Since the expansion is an even series of $\mu_B/T$, the square root is
expected to arise.
In order to calculate the zero order radius, we recall the results
at $\mu=0$ (not included in Eq.~(\ref{eq:pexpan1})). 
\begin{eqnarray}
\rho_o = \left(\frac{c_0}{c_2}\right)^{1/2} = \frac{T}{T_c}
\left[\frac{2}{9}
  \left(\frac{\epsilon_m(T)}{\epsilon_b(T,\mu_q=0)}+1\right)\right]^{1/2}. 
\end{eqnarray}
It is clear that $\rho_0$, in constast to the other radii, is $T$-dependent In
Fig.~\ref{Fig:5}, we find that close to $T_c$, $\rho_0=0.982$, $\rho_2=1.18$,
$\rho_4=1.82$ and $\rho_6=2.5$. These values are shown as short vertical
lines. The results agree very well with the available lattice
results~\cite{Allton:2003vx}; as $\rho_0\rightarrow 1$. The radius of
convergence approximately gives the lower bound for the critical endpoint
($\mu_q\geq T_c$). Obviously this value is higher than the recent lattice
results~\cite{Ejiri:2003dc, Fodor:2004nz}. In principle, 
one expects that due to the absence of critical behavior in HRGM, the
Taylor expansion, Eq.~(\ref{eq:pexpan1}), has an infinite convergence radius
for all temperatures. At $T_c$ and for the convergence radius $\rho=1$, one
expects that the hadron degrees of freedom are indistinguishable from
the degrees of freedom in QGP. On the other hand, the radii $\rho$ from 
lattice simulations are bounded from above by 
the location of phase transition. Therefore, it is expected that the radii
in Eq.~(\ref{eq:rhon}) stay close to unity near $T_c$. In
fact, this is the case for our low-order coefficients. The agreement
between our results and the lattice ones is convincing. There are no
published lattice results on the $6^{th}$-order radius of
divergence. Nevertheless, according to our expansion coefficients, we
expect that $\rho$ are steadily changing.

\section{\label{sec:8}Conclusions}

\begin{figure}[thb]
\includegraphics[width=10.cm]{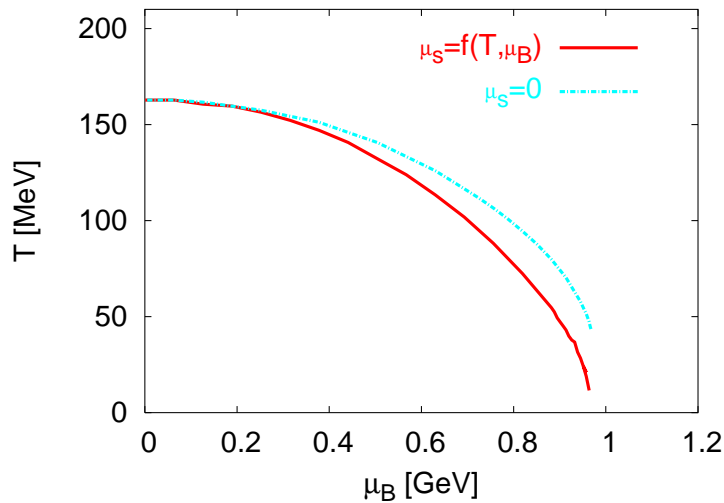}
\caption{\footnotesize Summary of our results on the QCD phase transition. In
  calculating both curves, we use all hadron resonances and suppose that the
  isospin chemical potential is vanishing. To obtain the top curve, we
  assigned $\mu_s$ to zero. $\mu_s$ is calculated in dependence on $T$
  and $\mu_B$. The results are given by the bottom curve. We note that both
  curves will cross the abscissa at the point of normal nuclear density. } 
\label{Fig:6}
\end{figure}

We used HRGM to draw up $T_c-\mu_B$ diagram. The transition
temperature $T_c$ from hadronic matter to QGP has been determined according
to a condition of constant energy density. Its value is taken from
lattice QCD simulations at zero chemical potential and assumed to remain
constant along the entire $\mu_B$-axis. We checked the influence of $s$ quark
chemical potential $\mu_s$ on $T_c$. For including $\mu_s$, we applied two
models. In the first one, we explicitly calculated $\mu_s$ in dependence on
$T$ and $\mu_B$ under the condition that the net strangeness vanishes. In the 
second one, we assigned, as the case in lattice QCD simulations, zero to
$\mu_s$ for all $T$ and $\mu_B$. The first case, $\mu_s=f(T,\mu_B)$, is of
great interest for heavy-ion collisions. Furthermore, under this
consideration, we expect that the strange quantum number is entirely
conserved. This is 
not the case in the second model. Nevertheless, with the last assignment,
the current lattice results are very well reproduced. On the other hand,
one can apply the second model for current and future heavy-ion
collisions. At BNL-RHIC and CERN-LHC energies, for instance, $\mu_B$ (and
consequently $\mu_s$) is very small.  We have shown that the
proper condition that guarantees vanishing strangeness in QGP is to set 
$\mu_s=\mu_q$. We did not check this explicitly. But it is obvious that
$T_c(\mu_B,\mu_s=0)$ quantitatively is not very much different from
$T_c(\mu_B,\mu_s(T,\mu_B))$ at small $\mu_B$.    

We have taken into consideration all hadron resonances given in particle
data booklet with masses up to $2\;$GeV. For a reliable comparison with
lattice, we have re-scaled the resonance masses to be comparable to the
quark masses used in lattice simulations. With excluding the strange
resonances, we compared our results with $n_f=2$ lattice results. With
including all resonances, we reproduced $n_f=2+1$ lattice results. We note
that increasing $\mu_B$ and $\mu_s$ leads to monotonic decrease in
$T_c$. The results from HRGM match very well with the lattice simulations,
especially within the $\mu_B$-range in which the lattice calculations are
most reliable ($\mu_B/T)\approx 3 T_c$. The agreement turns to be
excellent, when we take into consideration the truncations done in
calculating the thermodynamical quantity $\epsilon$.    

Besides this excellent agreement in the indirect calculation of $T_c$ via
constant energy density, we found that the analytical expression of 
$T_c(\mu)$ up to the second order of $(\mu/T)$ greatly
reproduced the curvatures calculated on lattice for different quark masses
and flavor numbers.  

Fig.~\ref{Fig:6} summarizes our conclusions. The QCD phase diagram is
plotted for a system including light as well as strange quarks. The Taylor
expansion of energy density is not truncated. The 
two curves represents our predictions, when it will be possible to perform
lattice simulations for physical masses and without the need to truncate
the Taylor expansion. As $T\rightarrow 0$, we note that both curves
will cross the abscissa at almost one point. It is obvious that this point
is corresponding to the normal nuclear density. The latter is related to
$\mu_B\sim 0.979\,$GeV. The nature of the phase transition at very low
temperatures does not lie within the scope of this work. However, there are
many indications that the transition at $T=0$ occurs according to
modification in the particle correlations. Changing the correlation leads
to quantum phenomena, like quantum entropy~\cite{Miller:2003ha,
  Miller:2003hh, Miller:2003ch, Miller:2004uc, Hamieh:2004ni,
  Miller:2004em}. We also note that by switching on $\mu_s$ an increase in
$T_c$ is expected. At small $\mu_B$, the two curves are coincide.   
 
\begin{acknowledgments}
We gratefully acknowledge the useful discussions with David~Blaschke,
Frithjof~Karsch, Berndt~M\"uller, Krzsyztof~Redlich, York~Schr\"oder and
Boris~Tomasik. 
\end{acknowledgments}


\begin{thebibliography}{30}
\expandafter\ifx\csname natexlab\endcsname\relax\def\natexlab#1{#1}\fi
\expandafter\ifx\csname bibnamefont\endcsname\relax
  \def\bibnamefont#1{#1}\fi
\expandafter\ifx\csname bibfnamefont\endcsname\relax
  \def\bibfnamefont#1{#1}\fi
\expandafter\ifx\csname citenamefont\endcsname\relax
  \def\citenamefont#1{#1}\fi
\expandafter\ifx\csname url\endcsname\relax
  \def\url#1{\texttt{#1}}\fi
\expandafter\ifx\csname urlprefix\endcsname\relax\def\urlprefix{URL }\fi
\providecommand{\bibinfo}[2]{#2}
\providecommand{\eprint}[2][]{\url{#2}}

\bibitem[{\citenamefont{Rajagopal and Wilczek}(2000)}]{Rajagopal:2000wf}
\bibinfo{author}{\bibfnamefont{K.}~\bibnamefont{Rajagopal}} \bibnamefont{and}
  \bibinfo{author}{\bibfnamefont{F.}~\bibnamefont{Wilczek}}
  (\bibinfo{year}{2000}), \eprint{hep-ph/0011333}.

\bibitem[{\citenamefont{Fodor and Katz}(2002)}]{Fodor:2001au}
\bibinfo{author}{\bibfnamefont{Z.}~\bibnamefont{Fodor}} \bibnamefont{and}
  \bibinfo{author}{\bibfnamefont{S.~D.} \bibnamefont{Katz}},
  \bibinfo{journal}{Phys. Lett.} \textbf{\bibinfo{volume}{B534}},
  \bibinfo{pages}{87} (\bibinfo{year}{2002}), \eprint{hep-lat/0104001}.

\bibitem[{\citenamefont{de~Forcrand and Philipsen}(2002)}]{deForcrand:2002ci}
\bibinfo{author}{\bibfnamefont{P.}~\bibnamefont{de~Forcrand}} \bibnamefont{and}
  \bibinfo{author}{\bibfnamefont{O.}~\bibnamefont{Philipsen}},
  \bibinfo{journal}{Nucl. Phys.} \textbf{\bibinfo{volume}{B642}},
  \bibinfo{pages}{290} (\bibinfo{year}{2002}), \eprint{hep-lat/0205016}.

\bibitem[{\citenamefont{Allton et~al.}(2002)}]{Allton:2002zi}
\bibinfo{author}{\bibfnamefont{C.~R.} \bibnamefont{Allton}}
  \bibnamefont{et~al.}, \bibinfo{journal}{Phys. Rev.}
  \textbf{\bibinfo{volume}{D66}}, \bibinfo{pages}{074507}
  (\bibinfo{year}{2002}), \eprint{hep-lat/0204010}.

\bibitem[{\citenamefont{D'Elia and Lombardo}(2003)}]{D'Elia:2002gd}
\bibinfo{author}{\bibfnamefont{M.}~\bibnamefont{D'Elia}} \bibnamefont{and}
  \bibinfo{author}{\bibfnamefont{M.-P.} \bibnamefont{Lombardo}},
  \bibinfo{journal}{Phys. Rev.} \textbf{\bibinfo{volume}{D67}},
  \bibinfo{pages}{014505} (\bibinfo{year}{2003}), \eprint{hep-lat/0209146}.

\bibitem[{\citenamefont{Gavai and Gupta}(2003)}]{Gavai:2003mf}
\bibinfo{author}{\bibfnamefont{R.~V.} \bibnamefont{Gavai}} \bibnamefont{and}
  \bibinfo{author}{\bibfnamefont{S.}~\bibnamefont{Gupta}},
  \bibinfo{journal}{Phys. Rev.} \textbf{\bibinfo{volume}{D68}},
  \bibinfo{pages}{034506} (\bibinfo{year}{2003}), \eprint{hep-lat/0303013}.

\bibitem[{\citenamefont{Karsch}(2002)}]{Karsch:2001cy}
\bibinfo{author}{\bibfnamefont{F.}~\bibnamefont{Karsch}},
  \bibinfo{journal}{Lect. Notes Phys.} \textbf{\bibinfo{volume}{583}},
  \bibinfo{pages}{209} (\bibinfo{year}{2002}), \eprint{hep-lat/0106019}.

\bibitem[{\citenamefont{Karsch et~al.}(2001)\citenamefont{Karsch, Laermann, and
  Peikert}}]{Karsch:2000kv}
\bibinfo{author}{\bibfnamefont{F.}~\bibnamefont{Karsch}},
  \bibinfo{author}{\bibfnamefont{E.}~\bibnamefont{Laermann}}, \bibnamefont{and}
  \bibinfo{author}{\bibfnamefont{A.}~\bibnamefont{Peikert}},
  \bibinfo{journal}{Nucl. Phys.} \textbf{\bibinfo{volume}{B605}},
  \bibinfo{pages}{579} (\bibinfo{year}{2001}), \eprint{hep-lat/0012023}.

\bibitem[{\citenamefont{Tawfik}(2004{\natexlab{a}})}]{Tawfik:2004vv}
\bibinfo{author}{\bibfnamefont{A.}~\bibnamefont{Tawfik}}
  (\bibinfo{year}{2004}{\natexlab{a}}), \eprint{hep-ph/0410329}.

\bibitem[{\citenamefont{Tawfik}(2004{\natexlab{b}})}]{Tawfik:2004ss}
\bibinfo{author}{\bibfnamefont{A.}~\bibnamefont{Tawfik}}
  (\bibinfo{year}{2004}{\natexlab{b}}), \eprint{hep-ph/0410392}.

\bibitem[{\citenamefont{Miller and Tawfik}(2003{\natexlab{a}})}]{Miller:2003ha}
\bibinfo{author}{\bibfnamefont{D.~E.} \bibnamefont{Miller}} \bibnamefont{and}
  \bibinfo{author}{\bibfnamefont{A.}~\bibnamefont{Tawfik}}
  (\bibinfo{year}{2003}{\natexlab{a}}), \eprint{hep-ph/0308192}.

\bibitem[{\citenamefont{Miller and Tawfik}(2003{\natexlab{b}})}]{Miller:2003hh}
\bibinfo{author}{\bibfnamefont{D.~E.} \bibnamefont{Miller}} \bibnamefont{and}
  \bibinfo{author}{\bibfnamefont{A.}~\bibnamefont{Tawfik}}
  (\bibinfo{year}{2003}{\natexlab{b}}), \eprint{hep-ph/0309139}.

\bibitem[{\citenamefont{Miller and Tawfik}(2003{\natexlab{c}})}]{Miller:2003ch}
\bibinfo{author}{\bibfnamefont{D.~E.} \bibnamefont{Miller}} \bibnamefont{and}
  \bibinfo{author}{\bibfnamefont{A.}~\bibnamefont{Tawfik}}
  (\bibinfo{year}{2003}{\natexlab{c}}), \eprint{hep-ph/0312368}.

\bibitem[{\citenamefont{Miller and Tawfik}(2004{\natexlab{a}})}]{Miller:2004uc}
\bibinfo{author}{\bibfnamefont{D.~E.} \bibnamefont{Miller}} \bibnamefont{and}
  \bibinfo{author}{\bibfnamefont{A.}~\bibnamefont{Tawfik}},
  \bibinfo{journal}{J. Phys.} \textbf{\bibinfo{volume}{G30}},
  \bibinfo{pages}{731} (\bibinfo{year}{2004}{\natexlab{a}}),
  \eprint{hep-ph/0402296}.

\bibitem[{\citenamefont{Hamieh and Tawfik}(2004)}]{Hamieh:2004ni}
\bibinfo{author}{\bibfnamefont{S.}~\bibnamefont{Hamieh}} \bibnamefont{and}
  \bibinfo{author}{\bibfnamefont{A.}~\bibnamefont{Tawfik}}
  (\bibinfo{year}{2004}), \eprint{hep-ph/0404246}.

\bibitem[{\citenamefont{Miller and Tawfik}(2004{\natexlab{b}})}]{Miller:2004em}
\bibinfo{author}{\bibfnamefont{D.~E.} \bibnamefont{Miller}} \bibnamefont{and}
  \bibinfo{author}{\bibfnamefont{A.}~\bibnamefont{Tawfik}},
  \bibinfo{journal}{Acta Phys. Polon.} \textbf{\bibinfo{volume}{B35}},
  \bibinfo{pages}{2165} (\bibinfo{year}{2004}{\natexlab{b}}),
  \eprint{hep-ph/0405175}.

\bibitem[{\citenamefont{Karsch et~al.}(2003{\natexlab{a}})\citenamefont{Karsch,
  Redlich, and Tawfik}}]{Karsch:2003zq}
\bibinfo{author}{\bibfnamefont{F.}~\bibnamefont{Karsch}},
  \bibinfo{author}{\bibfnamefont{K.}~\bibnamefont{Redlich}}, \bibnamefont{and}
  \bibinfo{author}{\bibfnamefont{A.}~\bibnamefont{Tawfik}},
  \bibinfo{journal}{Phys. Lett.} \textbf{\bibinfo{volume}{B571}},
  \bibinfo{pages}{67} (\bibinfo{year}{2003}{\natexlab{a}}),
  \eprint{hep-ph/0306208}.

\bibitem[{\citenamefont{Karsch et~al.}(2003{\natexlab{b}})\citenamefont{Karsch,
  Redlich, and Tawfik}}]{Karsch:2003vd}
\bibinfo{author}{\bibfnamefont{F.}~\bibnamefont{Karsch}},
  \bibinfo{author}{\bibfnamefont{K.}~\bibnamefont{Redlich}}, \bibnamefont{and}
  \bibinfo{author}{\bibfnamefont{A.}~\bibnamefont{Tawfik}},
  \bibinfo{journal}{Eur. Phys. J.} \textbf{\bibinfo{volume}{C29}},
  \bibinfo{pages}{549} (\bibinfo{year}{2003}{\natexlab{b}}),
  \eprint{hep-ph/0303108}.

\bibitem[{\citenamefont{Redlich et~al.}(2004)\citenamefont{Redlich, Karsch, and
  Tawfik}}]{Redlich:2004gp}
\bibinfo{author}{\bibfnamefont{K.}~\bibnamefont{Redlich}},
  \bibinfo{author}{\bibfnamefont{F.}~\bibnamefont{Karsch}}, \bibnamefont{and}
  \bibinfo{author}{\bibfnamefont{A.}~\bibnamefont{Tawfik}},
  \bibinfo{journal}{J. Phys.} \textbf{\bibinfo{volume}{G30}},
  \bibinfo{pages}{S1271} (\bibinfo{year}{2004}), \eprint{nucl-th/0404009}.

\bibitem[{\citenamefont{Toublan and Kogut}(2004)}]{Toublan:2004ks}
\bibinfo{author}{\bibfnamefont{D.}~\bibnamefont{Toublan}} \bibnamefont{and}
  \bibinfo{author}{\bibfnamefont{J.~B.} \bibnamefont{Kogut}}
  (\bibinfo{year}{2004}), \eprint{hep-ph/0409310}.

\bibitem[{\citenamefont{Hagedorn}(1965)}]{Hagedorn:1965st}
\bibinfo{author}{\bibfnamefont{R.}~\bibnamefont{Hagedorn}},
  \bibinfo{journal}{Nuovo Cim. Suppl.} \textbf{\bibinfo{volume}{3}},
  \bibinfo{pages}{147} (\bibinfo{year}{1965}).

\bibitem[{\citenamefont{Dashen et~al.}(1969)\citenamefont{Dashen, Ma, and
  Bernstein}}]{Dashen:1969mb}
\bibinfo{author}{\bibfnamefont{R.}~\bibnamefont{Dashen}},
  \bibinfo{author}{\bibfnamefont{S.-k.} \bibnamefont{Ma}}, \bibnamefont{and}
  \bibinfo{author}{\bibfnamefont{H.~J.} \bibnamefont{Bernstein}},
  \bibinfo{journal}{Phys. Rev.} \textbf{\bibinfo{volume}{187}},
  \bibinfo{pages}{345} (\bibinfo{year}{1969}).

\bibitem[{\citenamefont{Karsch}(2004)}]{Karsch:2004ti}
\bibinfo{author}{\bibfnamefont{F.}~\bibnamefont{Karsch}},
  \bibinfo{journal}{Prog. Theor. Phys. Suppl.} \textbf{\bibinfo{volume}{153}},
  \bibinfo{pages}{106} (\bibinfo{year}{2004}), \eprint{hep-lat/0401031}.

\bibitem[{\citenamefont{Ejiri et~al.}(2004)}]{Ejiri:2003dc}
\bibinfo{author}{\bibfnamefont{S.}~\bibnamefont{Ejiri}} \bibnamefont{et~al.},
  \bibinfo{journal}{Prog. Theor. Phys. Suppl.} \textbf{\bibinfo{volume}{153}},
  \bibinfo{pages}{118} (\bibinfo{year}{2004}), \eprint{hep-lat/0312006}.

\bibitem[{\citenamefont{Ali~Khan et~al.}(2001)}]{AliKhan:2001ek}
\bibinfo{author}{\bibfnamefont{A.}~\bibnamefont{Ali~Khan}} \bibnamefont{et~al.}
  (\bibinfo{collaboration}{CP-PACS}), \bibinfo{journal}{Phys. Rev.}
  \textbf{\bibinfo{volume}{D64}}, \bibinfo{pages}{074510}
  (\bibinfo{year}{2001}), \eprint{hep-lat/0103028}.

\bibitem[{\citenamefont{Karsch et~al.}(2004)}]{Karsch:2003va}
\bibinfo{author}{\bibfnamefont{F.}~\bibnamefont{Karsch}} \bibnamefont{et~al.},
  \bibinfo{journal}{Nucl. Phys. Proc. Suppl.} \textbf{\bibinfo{volume}{129}},
  \bibinfo{pages}{614} (\bibinfo{year}{2004}), \eprint{hep-lat/0309116}.

\bibitem[{\citenamefont{de~Forcrand and Philipsen}(2003)}]{deForcrand:2003hx}
\bibinfo{author}{\bibfnamefont{P.}~\bibnamefont{de~Forcrand}} \bibnamefont{and}
  \bibinfo{author}{\bibfnamefont{O.}~\bibnamefont{Philipsen}},
  \bibinfo{journal}{Nucl. Phys.} \textbf{\bibinfo{volume}{B673}},
  \bibinfo{pages}{170} (\bibinfo{year}{2003}), \eprint{hep-lat/0307020}.

\bibitem[{\citenamefont{Fodor}(2003)}]{Fodor:2002sd}
\bibinfo{author}{\bibfnamefont{Z.}~\bibnamefont{Fodor}},
  \bibinfo{journal}{Nucl. Phys.} \textbf{\bibinfo{volume}{A715}},
  \bibinfo{pages}{319} (\bibinfo{year}{2003}), \eprint{hep-lat/0209101}.

\bibitem[{\citenamefont{Fodor and Katz}(2004)}]{Fodor:2004nz}
\bibinfo{author}{\bibfnamefont{Z.}~\bibnamefont{Fodor}} \bibnamefont{and}
  \bibinfo{author}{\bibfnamefont{S.~D.} \bibnamefont{Katz}},
  \bibinfo{journal}{JHEP} \textbf{\bibinfo{volume}{04}}, \bibinfo{pages}{050}
  (\bibinfo{year}{2004}), \eprint{hep-lat/0402006}.

\bibitem[{\citenamefont{Allton et~al.}(2003)}]{Allton:2003vx}
\bibinfo{author}{\bibfnamefont{C.~R.} \bibnamefont{Allton}}
  \bibnamefont{et~al.}, \bibinfo{journal}{Phys. Rev.}
  \textbf{\bibinfo{volume}{D68}}, \bibinfo{pages}{014507}
  (\bibinfo{year}{2003}), \eprint{hep-lat/0305007}.

\end{thebibliography}

\end{document}